\pdfoutput=1
\documentclass[11pt]{article}
\usepackage[utf8]{inputenc}
\usepackage[top=50pt,bottom=50pt,left=68pt,right=66pt]{geometry}
\usepackage{amsmath}
\usepackage{hyperref}
\usepackage{amssymb}
\usepackage[space]{grffile}
\usepackage{graphicx}
\newcommand{\overbar}[1]{\mkern 1.5mu\overline{\mkern-1.5mu#1\mkern-1.5mu}\mkern 1.5mu}

\begin{document}
\thispagestyle{empty}

{\hbox to\hsize{
\vbox{\noindent Revised January 2019 \hfill IPMU18-0181 }}}

\noindent
\vskip2.0cm
\begin{center}

{\Large\bf Modified Born-Infeld-dilaton-axion coupling in supersymmetry}

\vglue.3in

Yermek Aldabergenov~${}^{a,b}$  and Sergei V. Ketov~${}^{c,d,e}$ 
\vglue.1in

${}^a$~Department of Physics, Faculty of Science, Chulalongkorn University,\\
Thanon Phayathai, Pathumwan, Bangkok 10330, Thailand\\
${}^b$~Institute of Experimental and Theoretical Physics, Al-Farabi Kazakh National University, \\
71 Al-Farabi Avenue, Almaty 050040, Kazakhstan \\
${}^c$~Department of Physics, Tokyo Metropolitan University, \\
Minami-ohsawa 1-1, Hachioji-shi, Tokyo 192-0397, Japan \\
${}^d$~Research School of High Energy Physics, Tomsk Polytechnic University,\\
30 Lenin Avenue, Tomsk 634050, Russian Federation \\
${}^e$~Kavli Institute for the Physics and Mathematics of the Universe (IPMU),
\\The University of Tokyo, Chiba 277-8568, Japan \\
\vglue.1in
aldyermek@gmail.com (YA), ketov@tmu.ac.jp (SVK)
\end{center}

\vglue.3in

\begin{center}
{\Large\bf Abstract}
\vglue.2in
\end{center}

We propose the supersymmetric extension of the modified Born-Infeld-axion-dilaton  non-linear electrodynamics that has confined static abelian solutions used for describing the electromagnetic confinement in the presence of axion and dilaton fields, and charged matter. The supersymmetric extension also has the non-trivial scalar potential that implies the upper bounds on the matter fields.

\newpage

\section{Introduction}
\vglue.1in

Born-Infeld (BI) theory \cite{Born:1934gh} is the profound extension of Maxwell electrodynamics, distinguished by its several theoretical features: (i) the Lagrangian as a density, (ii) Lorentz invariance, (iii) the resolution of the Coulomb singularity of the static electric field of a point-like charge, (iv) the upper limits on the values of electric and magnetic fields, (v) causal propagation of waves (no shock waves), (vi) the electric-magnetic self-duality, (vii) the origin in string theory as the low-energy effective theory of open strings
and D-branes --- see Refs.~\cite{Boillat:1970gw,Plebanski:1970zz,Fradkin:1985qd,Gaillard:1981rj,Ketov:1996bm,Grandi:1999dv,Olive:2000zv,Ketov:2001dq} for all these known aspects of the BI theory.

The supersymmetric extensions of the BI theory exhibit the additional highly non-trivial and unexpected features, such as (viii) the extra hidden (non-linearly) realized 
supersymmetries allowing (ix) the interpretation of the supersymmertic BI actions as the Maxwell-Goldstone actions associated with partial spontaneous supersymmetry breaking
\cite{Cecotti:1986gb,Bagger:1996wp,Rocek:1997hi,Ketov:1998ku,Tseytlin:1999dj} and
(x) adding Fayet-Iliopoulos (FI) terms \cite{Kuzenko:2009ym}.

The BI theory and its supersymmetric extensions are the very special and remarkable examples of the non-linear electrodynamics whose applicability is not limited to string theory. 
Moreover, for the sake of phenomenological applications, one may sacrifice some of the distinguished features (i-x) mentioned above. The phenomenological applications may also include the coupling to axion and dilaton fields to the non-linear electrodynamics, beyond the standard (minimal) coupling to the charged fields. 

One of such interesting applications was proposed in Ref.~\cite{Burton:2010gk}, where the
natural extension of the Born-Infeld action was used, which  preserves the properties (i)
and (ii) above. The authors of Ref.~\cite{Burton:2010gk} discovered the existence of {\it confined} solutions to the modified BI equations of motion, i.e. the fields with {\it finite} support in spacetime, in the presence of axion and dilaton fields. This feature is not shared
by the original BI theory and represents the electromagnetic {\it confinement} that may be of great interest to astrophysics and cosmology because it may imply the existence of new 
forms of matter (in the form of charged or uncharged stars invisible to us, like black holes). 

It is, therefore, of interest, to construct the supersymmetric extension of the modified BI
theory proposed in Ref.~\cite{Burton:2010gk}, and couple it to charged matter and a
dilaton-axion superfield. Our paper is devoted to this purpose.

The paper is organized as follows. In Sec.~2 we briefly review  the Born-infeld theory and
its $N=1$ supersymmetric extension in four spacetime dimensions. In Sec.~3 we also 
review the standard coupling of the BI theory to dilaton and axion, together with its supersymmetric extension. Secs.~2 and 3 provide the foundation for our new results given
in Secs.~4 and 5. Our conclusion is Sec.~6.
\vglue.2in

\section{Supersymmetric Born-Infeld theory}\label{BIS1}
\vglue.1in

The standard Born-Infeld (BI) Lagrangian reads \cite{Born:1934gh}
\begin{equation}
    {\cal L}_{\rm BI}=\frac{1}{b^2}\left(1-\sqrt{-{\rm det}(\eta_{mn}+bF_{mn})}\right)~,\quad 
    F_{mn}=\partial_mA_n-\partial_nA_m~,\label{BIbos}
\end{equation}
with the BI coupling constant $b$. In four spacetime dimensions, it takes the form
\begin{equation}
    {\cal L}_{\rm BI}=\frac{1}{b^2}\left(1-\sqrt{1+\frac{b^2}{2}F^2-\frac{b^4}{16}(F\Tilde{F})^2}\right)=-\frac{F^2}{4}+\frac{\frac{b^2}{16}[(F^2)^2+(F\Tilde{F})^2]}{1+\frac{b^2}{4}F^2+\sqrt{1+\frac{b^2}{2}F^2-\frac{b^4}{16}(F\Tilde{F})^2}}~,\label{BIbos2}
\end{equation}
where we have used the notation $\Tilde{F}^{mn}=\frac{1}{2}\epsilon^{mnkl}F_{kl}$,  
$F^2\equiv F_{mn}F^{mn}$ and $F\Tilde{F}\equiv\frac{1}{2}\epsilon^{mnkl}F_{mn}F_{kl}$ with Levi-Civita $\epsilon$.

The equations of motion (EOM) and Bianchi identities (BIs) of the BI theory can be written 
as  follows:
\begin{align}
    \partial^{n}G_{mn}=0~,\label{Fbianchi}\\
    \partial^{n}\Tilde{F}_{mn}=0~,\label{Geom}
\end{align}
respectively, where we have used the notation
\begin{equation}
    G_{mn}=-2\frac{\partial{\cal L}}{\partial F^{mn}}~.\label{GFrel}
\end{equation}
The EOM and BIs are invariant under the $SO(2)$ electric-magnetic duality rotations
\begin{align}
    F_{mn}\rightarrow\cos\gamma F_{mn}+\sin\gamma\Tilde{G}_{mn}~,\nonumber\\
    G_{mn}\rightarrow\cos\gamma G_{mn}+\sin\gamma\Tilde{F}_{mn}~,\label{emduality}
\end{align}
where we have used the notation $\Tilde{G}^{mn}=\frac{1}{2}\epsilon^{mnkl}G_{kl}$.
As regards a generic non-linear electrodynamics with the Lagrangian ${\cal L}(F)$, 
the condition of the electric-magnetic duality reads \cite{Gaillard:1981rj} 
\begin{equation}
    G\Tilde{G}=F\Tilde{F}~.\label{GGFF}
\end{equation}

The Lagrangian  \eqref{BIbos2} can be expanded as
\begin{equation}
    {\cal L}_{\rm BI}=-\frac{1}{4}F^2+\frac{b^2}{32}\left[(F^2)^2+(F\Tilde{F})^2\right]+\frac{b^4}{128}\left[(F^2)^3+F^2(F\Tilde{F})^2\right]+{\cal O}(b^6)~,
\end{equation}
whose leading term is the standard Lagrangian of Maxwell electrodynamics.

The form \eqref{BIbos2} of the BI theory is most convenient for its (rigid) supersymmetrization in superspace. In $N=1$ superspace, the supersymmetric BI theory  is described by the Lagrangian \cite{Deser:1980ck,Cecotti:1986gb}~\footnote{For a review of Born-Infeld theory and its supersymmetric extensions, see e.g., Refs.~\cite{Tseytlin:1999dj,Ketov:2001dq}.}
\begin{equation}
    {\cal L}_{\rm sBI}=\frac{1}{4}\left(\int d^2\theta W^2+{\rm h.c.}\right)+\frac{b^2}{4}\int d^4\theta\frac{W^2\overbar{W}^2}{1+\frac{b^2}{2}(\omega+\Bar{\omega})+\sqrt{1+b^2(\omega+\Bar{\omega})+\frac{b^4}{4}(\omega-\Bar{\omega})^2}}~,\label{SBI}
\end{equation}
in terms of the superfield strength $W^\alpha=-\frac{1}{4}\bar{D}^2D_\alpha V$ of the real gauge superfield $V$ with
$W^2\equiv W^\alpha W_\alpha$ and 
\begin{equation}
    \omega\equiv \frac{1}{8}D^2W^2=\frac{1}{4}(F^2-2D^2-iF\Tilde{F})+ \ldots~,
\end{equation}
where the dots stand for the fermionic and the higher order (in $\theta$) terms.

The Lagrangian \eqref{SBI} is invariant under the $U(1)$ gauge transformations
\begin{equation}
    V\rightarrow V-\Lambda-\overbar{\Lambda}
\end{equation}
with the chiral superfield gauge parameter $\Lambda$. In addition to the manifest $N=1$ supersymmetry (SUSY), the SUSY BI action is also known to be invariant under the second (non-linearly realized) supersymmetry \cite{Bagger:1996wp,Rocek:1997hi}.

The bosonic part of the Lagrangian \eqref{SBI} reads
\begin{equation}
    {\cal L}_{\rm sBI}=-\frac{A}{4}+\frac{\frac{b^2}{16}(A^2+B^2)}{1+\frac{b^2}{4}A+\sqrt{1+\frac{b^2}{2}A-\frac{b^4}{16}B^2}}=\frac{1}{b^2}\left(1-\sqrt{1+\frac{b^2}{2}A-\frac{b^4}{16}B^2}\right)~,\label{L2}
\end{equation}
where $A\equiv F^2-2D^2$ and $B\equiv F\Tilde{F}$.  The real auxiliary field $D$ can be eliminated by its equation of motion, $D=0$, then the resulting Lagrangian coincides with
Eqs.~\eqref{BIbos} and \eqref{BIbos2}.

The SUSY BI theory \eqref{SBI} can be further generalized by adding matter chiral superfields $\Phi_i$ (charged under the $U(1)$ gauge symmetry with charges $q_i$), together with a Fayet-Iliopoulos (FI) term:
\begin{equation}
    {\cal L}={\cal L}_{\rm sBI}+\int d^4\theta \left[ K(\overbar{\Phi}_ie^{q_iV}\Phi_i)
    +\xi V\right]~,
    \quad \Phi_i\to e^{q_i\Lambda}\Phi_i~,  \label{BIFIs}
\end{equation}
where we have introduced the arbitrary function $K$,~\footnote{We do not demand renormalizability.} and the real  constant FI parameter $\xi$. In terms of the field components, we find
\begin{equation}
    {\cal L}={\cal L}_{\rm sBI}-XD+ \ldots~,\label{BIFI}
\end{equation}
where we have introduced the notation
\begin{equation}
    X\equiv -\frac{1}{2}(K_V+\xi)\quad {\rm and} \quad K_V\equiv\frac{\partial K}{\partial V}|_{\theta=0}~,
\end{equation}
and the dots stand for the terms that are  irrelevant in the $D$ equation of motion.

When the matter fields are absent, $K=0$, the BI theory with the FI term \eqref{BIFIs} retains its hidden $N=2$ SUSY \cite{Kuzenko:2009ym}, although the FI term spontaneously breaks the linear $N=1$ SUSY. When the charged matter fields are present, the second non-linear SUSY is explicitly broken.

Given the Lagrangian \eqref{BIFI}, the EOM of $D$ reads
\begin{equation}
    D=X\sqrt{1+\frac{b^2}{2}(F^2-2D^2)-\frac{b^4}{16}(F\Tilde{F})^2}~,
\end{equation}
and its solution is given by
\begin{equation}
    D=\frac{X}{\sqrt{1+b^2X^2}}\sqrt{1+\frac{b^2}{2}F^2-\frac{b^4}{16}(F\Tilde{F})^2}~.
\end{equation}
Substituting the solution back into the Lagrangian yields
\begin{equation}
    {\cal L}=\frac{1}{b^2}\left(1-\sqrt{1+b^2X^2}\sqrt{1+\frac{b^2}{2}F^2-\frac{b^4}{16}(F\Tilde{F})^2}\right)~.
\end{equation}
Therefore, we obtain the scalar potential
\begin{equation}
    V_D=\frac{1}{b^2}\left(\sqrt{1+b^2X^2}-1\right)=\frac{1}{2}X^2-\frac{b^2}{8}X^4+\frac{b^4}{16}X^6+{\cal O}(b^6)~.
\end{equation}
\vglue.2in

\section{BI coupling to dilaton-axion, and its supersymmetrization}\label{BIS2}
\vglue.1in

The standard coupling of Born-Infeld theory to dilaton field $\phi$ and axion field $C$ reads~\footnote{The full theory also includes the dilaton and axion kinetic terms.}
\begin{align}
    \tilde{{\cal L}}_{\rm BI}&=\frac{1}{b^2}\left(1-\sqrt{-{\rm det}(\eta_{mn}+be^{\phi/2}F_{mn})}\right)+\frac{C}{4}F\Tilde{F}\nonumber\\&=\frac{1}{b^2}\left(1-\sqrt{1+\frac{b^2}{2}e^{-\phi}F^2-\frac{b^4}{16}e^{-2\phi}(F\Tilde{F})^2}\right)+\frac{C}{4}F\Tilde{F}~.\label{BIda}
\end{align}
In this case, the $SO(2)$ electromagnetic self-duality can be extended to the $SL(2,\mathbb{R})$ self-duality \cite{Gaillard:1981rj}~\footnote{In quantum theory (superstrings), the $SL(2,\mathbb{R})$ is broken to its discrete subgroup $SL(2,\mathbb{Z})$.}. After introducing the complex dilaton-axion field and its vacuum expectation value (VEV), 
\begin{equation}
    \tau\equiv C+ie^{-\phi}~,\quad 
    \tau_0=\frac{\Theta}{2\pi}+\frac{4\pi i}{e^2}~,\label{tauvev}
\end{equation}
respectively, where $\Theta$ is the vacuum theta-angle, and $e$ is the  $U(1)$ electric charge, the $SL(2,\mathbb{R})$ transformations read
\begin{equation}
    \tau\rightarrow\frac{a\tau+b}{c\tau+d}~, \quad {\rm with} \quad a,b,c,d\in\mathbb{R}~,
    \quad ad-bc=1~.
\end{equation}

The SUSY extension of the action \eqref{BIda} takes the form \cite{Ketov:2003gr}
\begin{equation}
    \tilde{{\cal L}}_{\rm sBI}=\frac{1}{4i}\left(\int d^2\theta\tau W^2-{\rm h.c.}\right)+\frac{b^2}{16}\int d^4\theta\frac{|\tau-\Bar{\tau}|^2W^2\overbar{W}^2}{1+\frac{b^2}{2}(\omega'+\Bar{\omega}')+\sqrt{1+b^2(\omega'+\Bar{\omega}')+\frac{b^4}{4}(\omega'-\Bar{\omega}')^2}}~,\label{sBIda}
\end{equation}
where we have used the notation
\begin{equation}
    \omega'\equiv\frac{1}{16i}(\tau-\Bar{\tau})D^2W^2=\frac{1}{4}e^{-\phi}(A-iB)+ \ldots~,
\end{equation}
and have promoted $\tau$ to the chiral superfield. The bosonic part of the Lagrangian \eqref{sBIda} reads
\begin{equation}
    \tilde{{\cal L}}_{\rm sBI}=\frac{1}{b^2}\left(1-\sqrt{1+\frac{b^2}{2}e^{-\phi}(F^2-2D^2)-\frac{b^4}{16}e^{-2\phi}(F\Tilde{F})^2}\right)+\frac{C}{4}F\Tilde{F}~.\label{csBIda}
\end{equation}
The auxiliary field $D$ is eliminated by its EOM, $D=0$, and the resulting Lagrangian coincides with \eqref{BIda}. In the parametrization \eqref{tauvev}, the Lagrangian should be rescaled as ${\cal L}\rightarrow \frac{e^2}{4\pi}{\cal L}$, in order to obtain the canonical kinetic term of $F_{mn}$.
\vglue.2in

\section{Modified BI theory with dilaton-axion-like couplings}\label{BIprimeS1}
\vglue.1in

As was argued in the Introduction, let us consider the modified coupling of the dilaton-axion field to BI theory, which was proposed in Ref.~\cite{Burton:2010gk}:
\begin{align}
    {\cal L}_{\rm BI'}&=\frac{1}{b^2}\left(1-\sqrt{-{\rm det}(\eta_{mn}+b\alpha F_{mn}+b\beta\Tilde{F}_{mn})}\right)\nonumber\\&=\frac{1}{b^2}\left(1-\sqrt{1+\frac{b^2}{2}\left(e^{-\phi}F^2-CF\Tilde{F}\right)-\frac{b^4}{16}\left(e^{-\phi}F\Tilde{F}+CF^2\right)^2}\right)~,\label{BIprime}
\end{align}
where the $\alpha$ and $\beta$ are related to the $\phi$ and $C$ as 
\begin{equation}
    \alpha^2-\beta^2=e^{-\phi}~,~~~-2\alpha\beta=C~.
\end{equation}
The Lagrangian \eqref{BIprime} is obtained from the original BI theory \eqref{BIbos} by the substitution $F_{mn}\rightarrow\alpha F_{mn}+\beta\Tilde{F}_{mn}$.~\footnote{As is clear from a comparison of Eqns. \eqref{BIda} and \eqref{BIprime} (see also Eqn. \eqref{sBIprime} below), the two theories differ in their $C$-dependence, but their leading ($b$-independent) terms coincide.}

We refer to the modified theory \eqref{BIprime} as the BI$'$ theory, and still associate the
fields $\phi$ and $C$  with the dilaton and axion, respectively, as in 
Ref.~\cite{Burton:2010gk}, because they regain their original meaning  in the weak coupling limit 
$b\to 0$. Indeed, in the BI$'$ theory, the shift symmetry $C\rightarrow C+const.$, defining the axion, is lost (together with the electromagnetic self-duality, since $G\Tilde{G}\neq F\Tilde{F}$), while $C$ directly affects the equations of motion for the electromagnetic field $A_{m}$. However, after expanding the action \eqref{BIprime},
\begin{multline}
    {\cal L}_{\rm BI'}=-\frac{1}{4}e^{-\phi}F^2+\frac{1}{4}CF\Tilde{F}+\frac{b^2}{32}(e^{-2\phi}+C^2)\left[(F^2)^2+(F\Tilde{F})^2\right]-\\-\frac{b^4}{128}(e^{-2\phi}+C^2)(e^{-\phi}F^2-CF\Tilde{F})\left[(F^2)^2+(F\Tilde{F})^2\right]+{\cal O}(b^6)~,
\end{multline}
in the leading order with respect to $b$, the BI$'$ theory coincides with the Maxwell theory coupled to the dilaton and axion. Hence, for the weak $F_{mn}$ field, the axionic shift symmetry approximately holds (as well as the approximate $SL(2,\mathbb{R})$ duality).

A supersymmetrization of the BI$'$ theory is straightforward and results in
\begin{equation}
      {\cal L}_{\rm sBI'}=\frac{1}{4i}\left(\int d^2\theta\tau W^2-{\rm h.c.}\right)+\frac{b^2}{4}\int d^4\theta\frac{|\tau|^2 W^2\overbar{W}^2}{1+\frac{b^2}{2}(\omega'+\Bar{\omega}')+\sqrt{1+b^2(\omega'+\Bar{\omega}')+\frac{b^4}{4}(\omega'-\Bar{\omega}')^2}}~,\label{sBIprime}
\end{equation}
where we have used the notation
\begin{equation}
    \omega'\equiv\frac{\tau}{8i}D^2W^2=-\frac{\tau}{4}(B+iA)+\ldots
\end{equation}
together with  $A\equiv F^2-2D^2$, $B\equiv F\Tilde{F}$, and $\tau=C+ie^{-\phi}+{\cal O}(\theta)$.

After expanding the Lagrangian in components, we find
\begin{gather}
    {\cal L}_{\rm sBI'}=-\frac{1}{4}(e^{-\phi}A-CB)+\frac{b^2}{16}\cdot\frac{|\tau|^2(A^2+B^2)}{1+\frac{b^2}{4}(e^{-\phi}A-CB)+\sqrt{1+\frac{b^2}{2}(e^{-\phi}A-CB)-\frac{b^4}{16}(e^{-\phi}B+CA)^2}}\nonumber\\=\frac{1}{b^2}\left(1-\sqrt{1+\frac{b^2}{2}(e^{-\phi}A-CB)-\frac{b^4}{16}(e^{-\phi}B+CA)^2}\right)\nonumber\\=\frac{1}{b^2}\left(1-\sqrt{1+\frac{b^2}{2}\left(e^{-\phi}F^2-2e^{-\phi}D^2-CF\Tilde{F}\right)-\frac{b^4}{16}\left(e^{-\phi}F\Tilde{F}+CF^2-2C D^2\right)^2}\right)~.\label{csBIprime}
\end{gather}
In the absence of matter fields and FI terms, the auxiliary field $D$ is eliminated by its 
EOM, $D=0$, and the resulting Lagrangian coincides with \eqref{BIprime}.

\subsection{Adding supersymmetric matter and FI term}\label{BIprimeS2}

Adding to the BI$'$ theory the charged matter and the FI term along the lines of Eqs.~\eqref{BIFIs} and \eqref{BIFI}, with
\begin{equation}
    {\cal L}'={\cal L}_{\rm sBI'}-XD+ \ldots~,\label{BI'FI}
\end{equation}
yields the EOM for $D$ in the form
\begin{multline}
    2b^2\psi D-\frac{b^4}{2}CD\left(\psi F\Tilde{F}+CF^2-2CD^2\right)=\\=2b^2X\sqrt{1+\frac{b^2}{2}\left(\psi F^2-2\psi D^2-CF\Tilde{F}\right)-\frac{b^4}{16}\left(\psi F\Tilde{F}+CF^2-2CD^2\right)^2}~,\label{EOM}
\end{multline}
where we have introduced the notation $\psi\equiv e^{-\phi}$. Since $D=0$ is no longer a valid solution for its EOM, this generates a scalar potential.

Let us study solutions to EOM of $D$. After setting $F_{mn}=0$, there are three of them (in terms of $D^2$),
\begin{gather}
    D^2_1=\frac{1}{3b^2C^2}\left[-4\psi-b^2X^2+\frac{(2\psi-b^2X^2)^2}{\lambda}+\lambda\right]~,\label{Dsq1}\\
    D^2_{2,3}=\frac{1}{3b^2C^2}\left[-4\psi-b^2X^2-\frac{\frac{1}{2}(1\pm i\sqrt{3})(2\psi-b^2X^2)^2}{\lambda}-\frac{1}{2}(1\mp i\sqrt{3})\lambda\right]~,\label{Dsq2}
\end{gather}
where we have used the notation
\begin{gather}
    \lambda\equiv\left(6b|X|\sqrt{3\sigma(\psi^2+C^2)}+\sigma+27(\psi^2+C^2)b^2X^2\right)^{\frac{1}{3}}~,\label{lambda}\\
    \sigma\equiv 8\psi^3+(15\psi^2+27C^2)(bX)^2+6\psi (bX)^4-(bX)^6~.\label{sigma}
\end{gather}

The two real solutions are $D_1=\pm\sqrt{D_1^2}$. For consistency, the sign of the square roots should be correlated with that of $X$, i.e. $\pm\sqrt{D^2_1}$ should correspond to $\pm|X|$, respectively.

Since $\sigma$ appears under the square root in \eqref{lambda}, when $\sigma<0$, the $D_1^2$ becomes complex, while  $D^2_{2,3}$ generically stay complex as well (however, for certain values of $X$, $\psi$, and $C$, the imaginary parts of $D_{1,2,3}$ may vanish). Furthermore, the $D^2_1$ has to be positive, i.e., the condition
\begin{equation}
    \frac{(2\psi-b^2X^2)^2}{\lambda}+\lambda-4\psi-b^2X^2\geq 0\label{lambdasigma}
\end{equation}
must hold. It can be easily checked that \eqref{lambdasigma} is always satisfied for 
$\sigma\geq 0$.

As is clear from the Lagrangian \eqref{csBIprime} or EOM \eqref{EOM}, the solution \eqref{Dsq1} does not guarantee that the expression under the square root is positive. This means we have to require that (for $F_{mn}=0$)
\begin{equation}
    1-b^2\psi D^2-\frac{b^4}{4}C^2 D^4\geq 0~,
\end{equation}
which yields (also demanding the $D^2$ to be positive)
\begin{equation}
    D^2\leq 2\frac{-\psi+\sqrt{\psi^2+C^2}}{b^2C^2}~.\label{Dsqcond}
\end{equation}

Substituting the solution \eqref{Dsq1} for $D^2$ leads to a complicated inequality involving $X$, $\psi$, and $C$.

We conclude that the supersymmetric BI$'$ theory leads to restrictions on the values of 
$\psi=e^{-\phi}$, $C$ and $X\equiv -\frac{1}{2}(K_V+\xi)$, in contrast to the standard supersymmertic BI theory, namely Eq.~\eqref{Dsqcond} together with the condition
\begin{equation}
\sigma\geq 0~.\label{sigmacond}
\end{equation}

\subsection{The case of a constant $\tau$}\label{consttau}

To get more insights, let us consider the particular case of $\tau=\tau_0$, i.e.
\begin{equation}
    \tau=\frac{\Theta}{2\pi}+\frac{4\pi i}{e^2}~.
\end{equation}

We find that the scalar potential of the supersymmetric BI$'$ theory has a restricted domain in terms of the values of $X$. In order to get numerical results, as an example, let us set 
$\Theta=2\pi$ and $e^2=4\pi$, so that $\tau=1+i$. We first derive the upper limit on $X$, using the inequality $\sigma\geq 0$ (with the definition \eqref{sigma}):
\begin{equation}
    -(bX)^6+6(bX)^4+42(bX)^2+8\geq 0~.\label{sigma1}
\end{equation}
The inequality \eqref{sigma1} is respected by~\footnote{Note that the BI parameter $b$ has the dimension $(\rm mass)^{-2}$ and can be written in terms of the BI mass scale as $b=M_{\rm BI}^{-2}$, whereas the FI parameter has the dimension of $(\rm mass)^{2}$.}
\begin{equation}
    |X|\equiv\frac{1}{2}|K_V+\xi|\leq \sqrt{5+3\sqrt{3}}b^{-1}\approx 3.19~b^{-1}~.\label{condX}
\end{equation}

\begin{figure}[t]
\caption{${\rm Im}D_1$ (vertical) versus $X$ (horizontal). For $|X|\leq \sqrt{5+3\sqrt{3}}$ (with $b=1$), the imaginary part of $D_1$ is zero, so the solution is valid in this region.}
\includegraphics[scale=0.6]{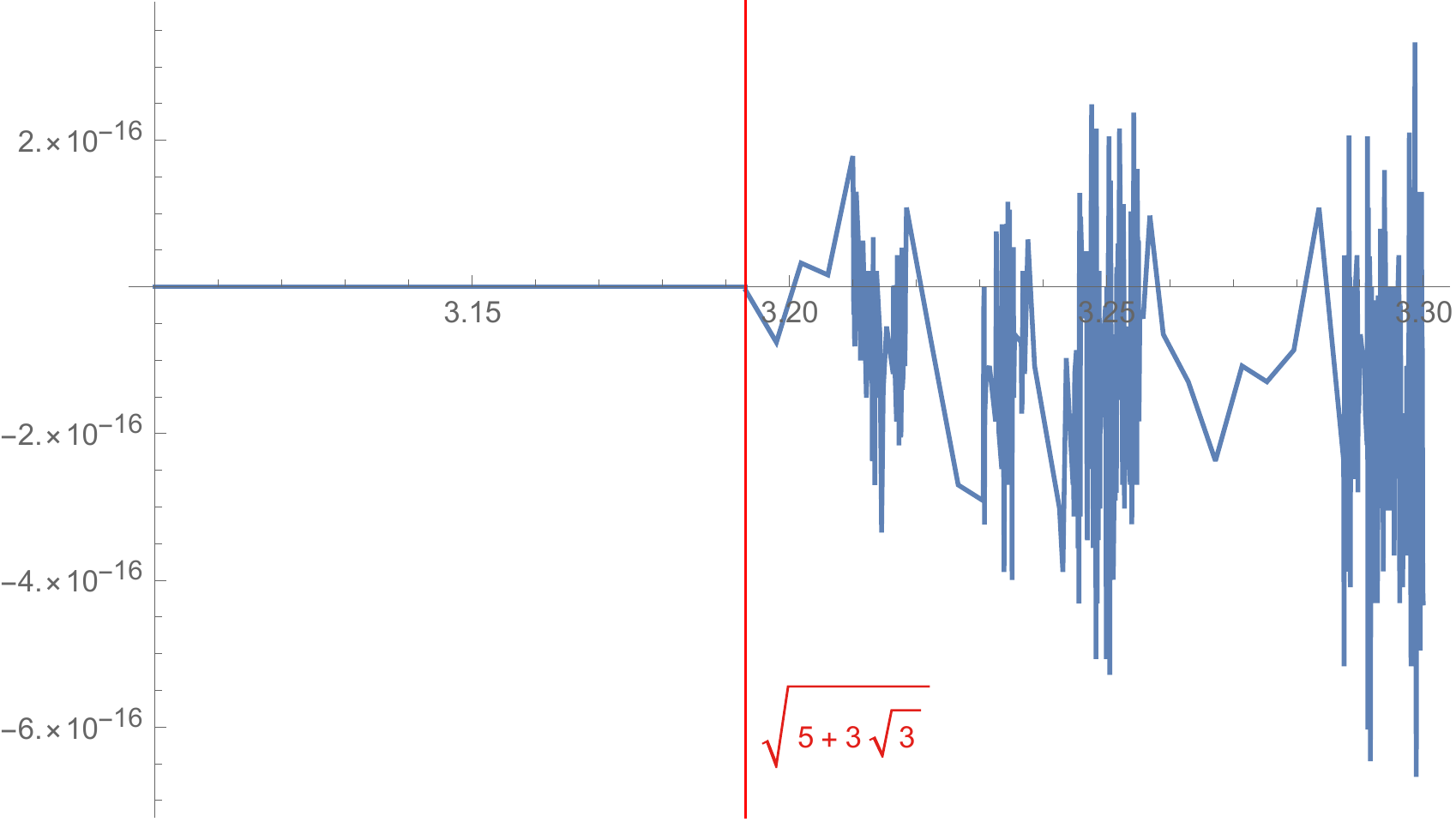}\label{F1}
\centering
\end{figure}

Figure \ref{F1} illustrates the solution $D_1$ acquiring the non-vanishing  imaginary part for $X>X_{\rm max}\equiv \sqrt{5+3\sqrt{3}}~b^{-1}$.

If we ignore the matter fields, $K=0$ , we arrive at the upper limit on the absolute value of the FI parameter:
\begin{equation}
    |\xi|\leq 2\sqrt{5+3\sqrt{3}}~b^{-1}~.\label{condxi}
\end{equation}
On the other hand, when the charged matter is present, the condition \eqref{condX} gives the upper limit on $K_V$ for a given $\xi$. For instance, with the canonical choice $K=\Phi e^{2V}\overbar{\Phi}$ and $K_V=2|\varphi|^2$, where $\varphi$ is the leading component of $\Phi$, the charged scalar $\varphi$ is bounded from above. Setting $\xi=0$ for simplicity, we find
\begin{equation}
    |\varphi|^2\leq \sqrt{5+3\sqrt{3}}~b^{-1}~.
\end{equation}
Note that the upper limit of $|X|$ is of the same order ($1/b$) as that for $F_{mn}$.

Next, let us consider the condition \eqref{Dsqcond}. When $\psi=C=1$, it takes the form
\begin{equation}
    |D|\leq\sqrt{2\sqrt{2}-2}~b^{-1}~.\label{Dsqcond2}
\end{equation}
After a substitution of the solution $\eqref{Dsq1}$ into \eqref{Dsqcond2}, the resulting inequality can be numerically solved for $X$, yielding the conditions
\begin{equation}
    |X|\lessapprox 2.02~b^{-1} \quad {\rm and} \quad |X|\gtrapprox 3.04~b^{-1}~.
\end{equation}

\section{The scalar potential}\label{scalarpot}

The scalar potential of the supersymmetric matter-coupled BI$'$ theory \eqref{BI'FI} can be obtained by using the solutions \eqref{Dsq1} for $D$ and setting $F_{mn}=0$. The exact potential is very complicated, and seems impossible to obtain its perturbative expansion in terms of $b$, because the solution to $D$ has the overall factor of $b^{-1}$.

Though the exact features of the potential depend on the choice of $X$, it is nevertheless possible to get some general results. For instance, the potential is always non-negative, and its minimum is at $X=0\rightarrow D=0$, if it exists, with unbroken SUSY and Minkowski vacuum. This is similar to the ordinary $D$-term potentials proportional to $D^2$. With the 
positive FI term and no matter, the minimum is of the de Sitter type with the $D$-term SUSY breaking.

The main difference against the ordinary $D$-term potentials is the existence of restrictions on the values of $X\equiv\frac{1}{2}(K_V+\xi)$ for given values of $\psi=e^{-\phi}$ and $C$. These restrictions arise from the conditions \eqref{sigmacond} and \eqref{Dsqcond2} for general $\psi$ and $C$. If we set $\psi=C=1$, we obtain the specific results:
\begin{equation}
    |X|\lessapprox 2.02~,~~~3.04\lessapprox |X|\lessapprox 3.19,~
\end{equation}
in units of $b=1$.

\begin{figure}[t]
\caption{The scalar potential (vertical) versus $X$ (horizontal). The dashed red lines mark the values $X\approx 2.02$ and $X\approx 3.04$, while solid red line stands at $X\approx 3.19$, the upper limit on $X$.}
\includegraphics[scale=0.6]{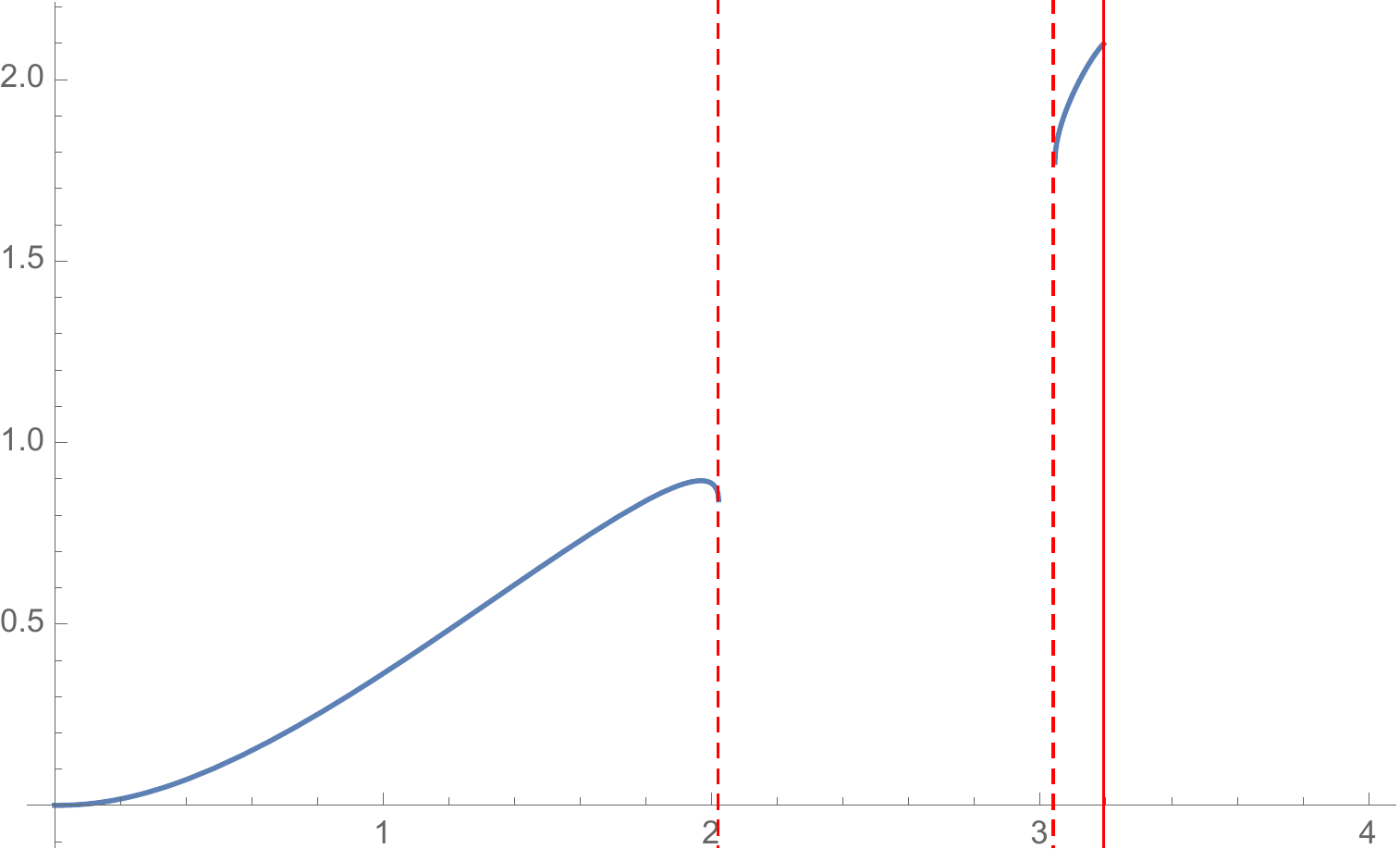}\label{F2}
\centering
\end{figure}

We provide the plot of the scalar potential as the function of $X$ in Figure \ref{F2}. For negative values of $X$, the negative root of $D^2_1$ should be used as the solution to $D$, so that the potential is symmetric under $X\rightarrow -X$.

\section{Conclusion}\label{concl}

Our main results are given in Secs. 4 and 5. 

It is of further interest to explore whether the confining mechanism of 
Ref.~\cite{Burton:2010gk} also exists in non-abelian extensions of the (modified) 
Born-Infeld theory and/or their supersymmetric extensions, in the presence of the dilaton-axion superfield. For example, a non-abelian (and non-supersymmetric) Born-Infeld (NBI) theory with a $\Theta$-term, having a similar structure to the BI$'$ theory, was considered in Ref.~\cite{Grandi:1999dv}, where it was found that this NBI theory (together with Higgs sector) admits monopole solutions, and has Witten's effect (i.e. a shift of the electric charge of the monopole induced by the presence of $\Theta$-term \cite{Witten:1979ey}), like that in the usual non-abelian gauge theory. Unfortunately, a non-abelian extension of the BI theory and, hence, a supersymmetric NBI theory is not unique \cite{Ketov:2001dq}.

\section*{Acknowledgements}

Y.A. was supported by the CUniverse research promotion project of Chulalongkorn University under the grant reference CUAASC, and the Ministry of Education and Science of the Republic of Kazakhstan under the grant reference AP05133630. S.V.K. was supported by a Grant-in-Aid of the Japanese Society for Promotion of Science (JSPS) under No.~26400252, the Competitiveness Enhancement Program of Tomsk Polytechnic University in Russia, and the World Premier International Research Center Initiative (WPI Initiative), MEXT, Japan. 

\bibliography{Bibliography.bib}{}
\bibliographystyle{utphys.bst}

\end{document}